\documentclass[sigconf,authorversion]{acmart}
\AtBeginDocument{%
  \providecommand\BibTeX{{%
    \normalfont B\kern-0.5em{\scshape i\kern-0.25em b}\kern-0.8em\TeX}}}

\acmConference[]{}{}{}

\usepackage{amsmath}
\usepackage{xspace}
\usepackage{balance}
\usepackage{enumitem}
\usepackage{ifthen}
\usepackage[skip=2pt]{caption}
\usepackage{subcaption}
\usepackage{hyperref}
\usepackage{multirow}
\usepackage{graphicx}
\usepackage{framed}

\setcopyright{none}
\renewcommand\footnotetextcopyrightpermission[1]{} 
{\endMakeFramed}

\newboolean{showcomments}
\setboolean{showcomments}{true}

\ifthenelse{\boolean{showcomments}}
{\newcommand{\nb}[2]{
		\fbox{\bfseries\sffamily\scriptsize#1}
		{\sf\small$\blacktriangleright$\textit{#2}$\blacktriangleleft$}
	}
}
{\newcommand{\nb}[2]{}
}

\newcommand{\ie}{\textit{i.e.,}\xspace}
\newcommand{\eg}{\textit{e.g.,}\xspace}
\newcommand{\etc}{\textit{etc.}\xspace}
\newcommand{\etal}{\textit{et al.}\xspace}

\newcommand{\prof}{\texttt{PROF}\xspace}
\newcommand{\stud}{\texttt{STUD}\xspace}
\newcommand{\torino}{\texttt{NI}\xspace}
\newcommand{\napoli}{\texttt{SI}\xspace}
\newcommand{\roma}{\texttt{CI}\xspace}

\newcommand{\task}{$TaskCompletion$\xspace}
\newcommand{\speed}{$Speed$\xspace}

\acmSubmissionID{}

\begin{document}

\title{Further Evidence on a Controversial Topic about Human-Based Experiments: Professionals vs. Students}

\author{Simone Romano}
\affiliation{%
 \institution{University of Salerno}\city{Fisciano (SA)}\country{Italy}
}
\email{siromano@unisa.it}

\author{Francesco Paolo	Sferratore}
\affiliation{%
  \institution{University of Salerno}
  \city{Fisciano (SA)}
  \country{Italy}
}
\email{f.sferratore@studenti.unisa.it}

\author{Giuseppe Scanniello}
\affiliation{%
  \institution{University of Salerno}
  \city{Fisciano (SA)}
  \country{Italy}
}
\email{gscanniello@unisa.it}


\begin{abstract} 

Most Software Engineering (SE) human-based controlled experiments rely on students as participants, raising concerns about their external validity. Specifically, the realism of results obtained from students and their applicability to the software industry remains in question.  
In this short paper, we bring further evidence on this controversial point. To do so, we compare 62 students and 42 software professionals on a bug-fixing task on the same Java program. The students were enrolled in a Bachelor's program in Computer Science, while the professionals were employed by two multinational companies (for one of them, the professionals were from two offices). Some variations in the experimental settings of the two groups (students and professionals) were present. 
For instance, the experimental environment of the experiment with professionals was more realistic; \ie they faced some stress factors such as interruptions during the bug-fixing task. 
Considering the differences between the two groups of participants,
the gathered data show that the students outperformed the professionals in fixing bugs. 
This diverges to some extent from past empirical evidence. Rather than presenting definitive conclusions, our results aim to catalyze the discussion on the use of students in experiments and pave the way for future investigations. Specifically, our results encourage us to examine the complex factors influencing SE tasks, making experiments as more realistic as possible.

\end{abstract}


\maketitle

\section{Introduction} \label{sec:Introduction}
Empirical research plays an important role in advancing Software Engineering (SE), providing evidence-based knowledge on the effectiveness of practices, techniques, and tools. Controlled experiments are one of the most widely used empirical research strategies in SE~\cite{Salman:2015,Guevara-Vega:2021}. Controlled experiments can be \textit{human-} or \textit{technology-based}. In human-based experiments, human participants receive treatments, while in technology-based experiments, different technical treatments are applied without the involvement of human participants~\cite{Wohlin:2012}. A significant challenge in human-based controlled experiments (from here on, controlled experiments or simply experiments) is the recruitment of participants, which has pushed researchers to use students in Computer Science (CS) or Computer Engineering (CE), instead of software professionals, as the participants~\cite{Salman:2015}. Using students could raise some concerns about the generalizability of the findings to the population of software professionals~\cite{Salman:2015}. In other words, the generalizability of such controlled experiments and the validity of findings derived from academic settings could cast doubt on how transferable these results are to real-world software development~\cite{Runeson2003}.

The comparison of students and professionals as participants in experiments has long been debated due to questions about their realism~\cite{Sjoberg:ESEM2002,Sjoberg:TSE2005,FalessiEtAl:EMSE2018,Feldt:2018}. According to Sj{\o}berg~\etal~\cite{Sjoberg:ESEM2002}, experiments with students are unrealistic in terms of environment, tasks, and clearly for the kind of participants. When experiments lack realism, the findings are only valid in a specified experimental situation, making the results less significant for both applied and theoretical research~\cite{Sjoberg:ESEM2002}. 
Moreover, students are usually motivated by grades, course credits, or learning opportunities. This could represent another limitation, as motivation can make a difference when comparing students and professionals~\cite{FalessiEtAl:EMSE2018}. Nevertheless, it is recognized that the participation of students in experiments has several advantages, such as a more homogeneous group of participants, the availability of a large number of participants, and the chance to test experimental design and initial hypotheses~\cite{Sjoberg:TSE2005,VerelstEMSE2005,Carver:2003}. Moreover, past studies have shown that students and professionals perform similarly when applying a technology that is new to both of them during
experimentation or when students are adequately trained for the task at hand~\cite{Salman:2015,Host:2000}. 
Finally, professionals, besides being difficult to recruit, have heterogeneous backgrounds. And, conducting experiments with professionals is costly (compensating professionals for their time can make studies expensive). 

In this short paper, we contribute to the ongoing debate on the realism of results from experiments with students, compared with those with professionals, and their applicability to the software industry. To do so, we compare 62 students and 42 professionals in a bug-fixing task on the same Java program. The students were enrolled in a Bachelor's program in CS. The professionals, instead, worked for two multinational companies; for one of them, the professionals were from two different offices. The experimental environments differed between the two groups. For instance, the professionals faced some stress factors such as interruptions during the bug-fixing task, creating a setting that more closely mirrored real-world industry conditions. The experimental data indicated that the students performed better than the professionals when fixing bugs. Rather than drawing definitive conclusions, our findings catalyze further discussion and investigation. In particular, our findings encourage us to examine the complex factors influencing SE tasks, making controlled experiments (and their environmental conditions) as realistic as possible.

\section{Motivation and Related Work} \label{sec:relatedWork}
Realism in experiments is crucial and often demands greater resources to ensure their validity~\cite{Sjoberg:ESEM2002}. Mundane realism, which refers to how closely an experiment mirrors real-world scenarios, directly impacts our ability to generalize the findings to industrial practice. Achieving realism in experimental tasks, participant selection, and environmental conditions presents a significant challenge. A widely debated criticism of experimental research is the frequent reliance on students as participants, which may limit the generalizability of results to professional settings~\cite{Salman:2015,Runeson2003,FalessiEtAl:EMSE2018,Sjoberg:ESEM2002}. In other words, how well do students represent professionals as experimental subjects? 
In this respect, H\"{o}st \etal~\cite{Host:2000} compared professionals (\ie project managers) and graduate students on a non-trivial SE task performed individually. The participants were asked to judge the effect of ten factors (\eg competence, product complexity, \etc) on the lead time of projects using the {Analytic Hierarchy Process} method. Both the project managers and students were asked to characterize projects for each of the ten factors. The two groups were then compared according to the actual effort and lead time. Only minor differences were observed between the performance of project managers and students. It was concluded that students, if adequately trained, can be used instead of professionals. Counsell~\cite{Counsell:2008} analyzed a dataset
containing the ratings of students and professionals
about class cohesion. No substantial difference was observed between the ratings of these two groups. Runeson~\cite{Runeson2003} hypothesized, within the context of the {Personal Software Process}, that graduate students perform similarly to professionals, while the performance of freshmen students differs. To test this, the author conducted a quantitative analysis comparing freshmen and graduate students, while also examining improvement trends against industry data. The hypothesis was not fully supported across all the measured dimensions, such as lines of code, defect count, and productivity. While clear differences emerged between freshmen and graduate students, the available data was insufficient to determine whether the performance of graduate students aligned with that of professionals.
Salman~\etal~\cite{Salman:2015} compared graduate students and professionals to understand from a quantitative perspective how well students represent professionals as experimental subjects in SE research. The context was that of two experiments on {Test-driven Development}, one conducted with students in an academic setting and one conducted with professionals in an industry setting. The authors measured the code quality of several tasks implemented by both the students and professionals and then checked whether they performed similarly in terms of code quality metrics. Except for minor differences, neither of the groups was better than the other. It was concluded that students and professionals perform similarly when applying a new technology. Falessi~\etal~\cite{FalessiEtAl:EMSE2018} performed a focus group, followed by a survey, to gain insight into the pros and cons of using students in SE (controlled) experiments. First, the authors elicited the opinions of 65 empirical SE experts on this matter. Afterward, they derived 14 statements and asked 27 empirical SE experts to provide their level of agreement with these statements. It was concluded that employing students as participants remains a justifiable simplification of reality in laboratory settings. This approach effectively contributes to advancing SE theories and technologies. However, as with any other design choice, careful consideration is required throughout the study planning, execution, analysis, and reporting. The critical factor is understanding which segment of the developer population the participants represent.
Carver~\etal~\cite{Carver:2003} identified four key figures involved in SE empirical studies: researchers, students, instructors, and industry. Afterward, they discussed the costs and benefits of using students as participants with respect to these figures. 
The authors also provided recommendations for conducting empirical studies (including controlled experiments) with students in SE. 
Finally, 
Sj{\o}berg~\etal~\cite{Sjoberg:TSE2005}, in their literature study, quantitatively characterized
experiments published in high-quality venues (from 1993 to 2002) in terms of their participants (\eg students vs. professionals), tasks (\eg type or duration), and environments (\eg location or development tools). The authors observed that an important issue concerning SE research is its relevance to the industry (\eg professionals are scarcely used as participants or experimental tasks are often not~realistic). 

Our study distinguishes itself from previous work in several ways. Unlike past studies, except for a few~\cite{Salman:2015,Host:2000,Runeson2003,Counsell:2008,PorterVotta1998}, we performed a large quantitative comparison between professionals and students (104 participants in total), thereby offering a robust empirical basis for understanding how these two groups differ. This approach not only strengthens the validity of our findings but also provides deeper insights into the nuances of performance in real-world settings. Moreover, our research is the only one to focus on a task that professionals have identified as particularly stressful compared to other SE tasks~\cite{RomanoCGVCSPROFES2024}. By selecting a task that mirrors the real pressures encountered in practice, we enhance the practical relevance of our study and underscore the value of investigating performance under conditions that truly matter to the industry. The considered sample further enriches our study by incorporating professionals from two multinational companies, one of them contributes with participants from two offices. This diversity in organizational and geographical contexts contributes significantly to the external validity of our work, offering a more comprehensive view of professional performance in varied settings. Perhaps most strikingly, our main finding challenges the conventional wisdom on the use of students in controlled experiments since we found that students outperformed professionals. This outcome, unexpected in light of the common behavior observed in past experiments, adds an important dimension to the ongoing debate regarding the realism of results from experiments with students.

\section{Methodology} \label{sec:studyDesign}
To contribute to the debate on the use of students versus professionals in controller experiments, we investigated the following high-level Research Question (RQ):
\textit{Do students and software professionals perform similarly when fixing bugs in Java programs?}
To answer this RQ, we used the data from two empirical studies: a \textit{controlled experiment}, not yet published, conducted in an industrial setting with professionals within the context of the MOOD project~\cite{MOOD:2024}; and a \textit{prospective observational study}, already published and with data and experimental material publicly available~\cite{Romano:2024:PROFES}, conducted in an academic setting with students. From here onward, we refer to these studies as {\prof} and {\stud}, respectively. Both groups were asked to execute a bug-fixing task on the same Java program. We considered such a kind of task because past work showed that professionals deem bug fixing as particularly stressful~\cite{RomanoCGVCSPROFES2024}. 

\begin{figure}[t]
\includegraphics[width=0.7\linewidth]{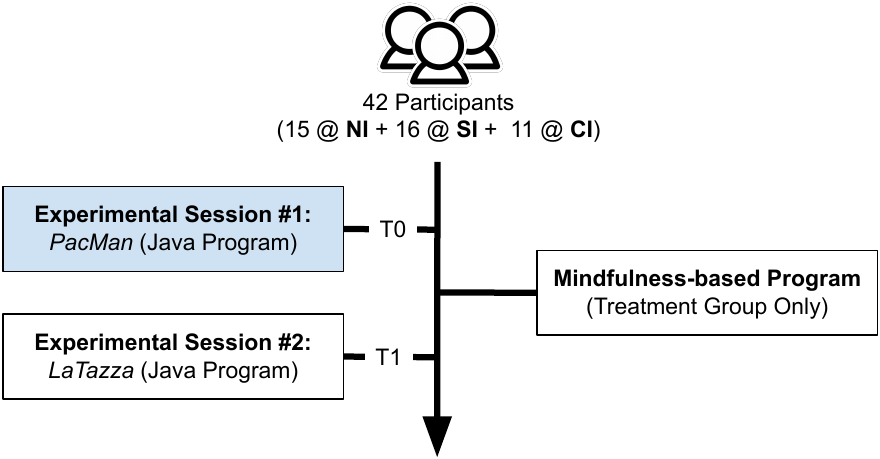}
\caption{Summary of \prof.}\label{fig:prof}
\vspace{-0.4cm}
\end{figure}

\begin{figure}[t]
\includegraphics[width=0.7\linewidth]{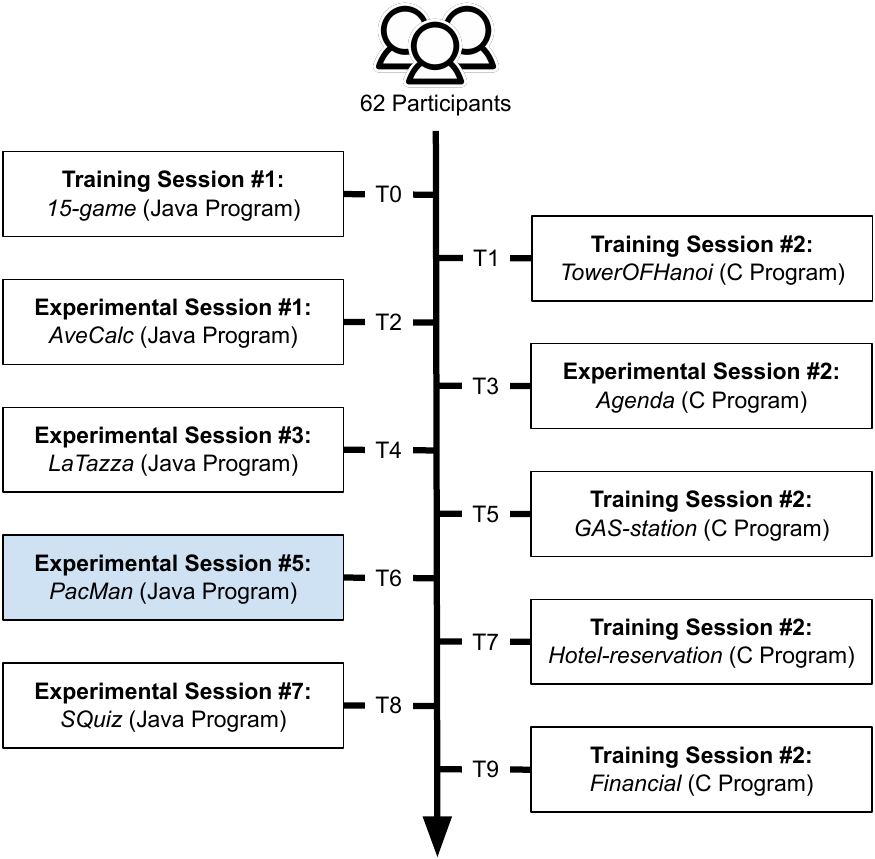}
\caption{Summary of \stud.}\label{fig:stud}
\vspace{-0.5cm}
\end{figure}

\textbf{\prof.} The controlled experiment aimed to estimate the benefits of a mindfulness-based program when fixing bugs. A graphical summary of this experiment is shown in Figure~\ref{fig:prof}. The experiment involved 42 software professionals working for two multinational companies operating in the Italian software industry. The experiment was multi-site since it took place at: two offices, located in Northern and Southern Italy, of the same multinational company; and one office of another multinational company located in Central Italy. From here onward, we also refer to these three sites as {\torino} (Northern Italy), {\napoli} (Southern Italy), and {\roma} (Central Italy), respectively. The distribution of the participants across the three sites was the following: 15 at \torino, 16 at \napoli, and 11 at \roma. Regardless of the site, the participants took part in a first experimental session, at the time \texttt{T0} (\ie before the attendance to the mindfulness program), in which they were asked to fulfill a bug-fixing task on a Java program, \textit{PacMan}. Later, a group of participants (\ie the treatment group) underwent a mindfulness-based program, while the remaining participants (\ie the control group) did not. 
After the mindfulness-based program, all the professionals were asked to fulfill another bug-fixing task at the time \texttt{T1}. 
To answer RQ, we used the data from the first experimental session (at \texttt{T0}) of each site. Namely, we relied our analysis on the data we gathered before the administration of the treatment to avoid any effect thereof.

\textbf{\stud.} The prospective observational study aimed to discover possible relationships between personality traits and performance in fixing bugs---as in any prospective observational study, no treatment was received by the participants~\cite{Saarimaki:2020}. A graphical summary of this study is shown in Figure~\ref{fig:stud}. The study involved 62 undergraduates in CS who took part in eight experimental sessions. At each time (from \texttt{T0} to \texttt{T9}), the students were asked to fix bugs in a C or Java program under controlled conditions in a laboratory at the University of Basilicata, Italy. Before the experimental sessions took place, the students were trained to fix bugs since, unlike professionals, they were not accustomed to doing so. Specifically, the students attended two training sessions at \texttt{{T0}} and \texttt{{T1}}: one on fixing bugs in Java code and one on fixing bugs in C code. In the first part of the training sessions, an educator showed how to fix some sample bugs, given the corresponding bug reports, in C or Java code; different strategies on how to identify and fix these sample bugs were covered and discussed with the students. In the second part of the training sessions, the participants experimented with bug fixing in C or Java code alone. 
In this paper, we used the data from the fifth experimental session where students worked on \textit{PacMan}---the same Java program as the first experimental session of \prof.

\subsection{Participants}
\hspace{\parindent}\ignorespaces
\textbf{\prof.} The participation was voluntary and not rewarded. The participants were recruited through an internal notice of the company they worked for. The internal notice described the aim of the research and the type of activities proposed. 
The average seniority of the 42 participants as a software professional was equal to 6.67 years for \torino, 10.06 years for \napoli, and years 2.45 for \roma. As for their average self-rated experience in Java (over a 1-5 rating scale where 1 means ``inexpert''), it was equal to 3.4, 3.7, and 3.9, respectively. Their average self-rated experience in bug fixing (over a 1-5 rating scale where 1 means ``inexpert'') was, instead, equal to 3.27, 3.63, and 3.09, respectively. 
In terms of their (highest) education degree, the number of master's graduates in CS or CE was equal to one for \torino (7\%), four for \napoli (25\%), and five for \roma (45\%).

\textbf{\stud.} The participants in this study were sampled by convenience among the students attending an SE course at the University of Basilicata (Italy). This course was scheduled in the first semester of the third (and last) year of the CS program. Participation was voluntary but rewarded with two bonus points on the final mark of the course. 
The 62 participants declared to be all skilled at programming in Java, while they were inexpert at fixing bugs---this was why we specifically trained them on bug fixing. 

\subsection{Experimental Task and Instrumentation}
For both \prof and \stud, we analyzed (as mentioned above) the data from a bug-fixing task on the same program, \ie~\textit{PacMan}. It is a Java implementation of the popular \textit{Pac-Man} game. The program consisted of four Java files and 1.4 KLOC (\ie thousands of lines of code, excluding comments and blank lines). The bugs in \textit{PacMan} were seeded by applying mutation operators and independent of one another. The use of mutation operators was to simulate errors that developers usually make~\cite{Jia:2011}.
Regardless of the study, the participants were given an ordered list of bug reports, each associated with a bug to be fixed, and had to tackle the bugs one at a time in the provided order. The participants were allowed to skip a bug when they could not fix it. Once a bug was fixed or skipped, the participants could not go back. The task was individual and the participants were allowed to use any means to identify and fix the bugs but not AI (Artificial Intelligence). Namely, the participants could use any non-AI feature of their IDE (Integrated Development Environment) such as the debugger, and online resources such as \textit{Stack Overflow}. The participants were monitored by the experimenters while carrying out the task.

\textbf{\prof.} Regardless of the site, the participants were asked to fix six bugs in a fixed time of 45 minutes. During the execution of the task, there were three interruptions, each simulating a message from the customer or project manager. The messages asked if the participant had finished fixing bugs and, if not, the number of leftover bugs. This design choice was to create a setting that more closely mirrored real-world industry conditions, where developers face time pressure and stress factor~\cite{RomanoCGVCSPROFES2024}. The participants used the IDE they preferred while fixing bugs. The bug reports were available on the issue tracker of the project on \textit{GitHub}. After fixing a bug, the participants had to perform a \textit{Git} commit to store the corresponding patch in the local \textit{Git} repository.  That is, we used the local \textit{Git} repository of each participant to collect his/her patches. 

\begin{table}[t]
\caption{Differences between \prof and \stud.}
\label{tab:diff}
\scriptsize
\begin{tabular}{@{}lp{2.8cm}p{2.8cm}@{}}
\toprule
Characteristic & \prof                                & \stud                          \\ \midrule
\#Bugs to be fixed         & 6                                   & 4    \\ 
Time allotted & 45 minutes, 3 interruptions & 40 minutes, no interruption   \\ 
Used IDE                   & Chosen by the participants           & \textit{NetBeans} \\ 
Sharing of bug report & \textit{GitHub}'s issue tracker & \textit{Google Forms}' questionnaire \\ 
Collection of patches & Local \textit{Git} repository, after committing patches & \textit{Google Forms}' questionnaire, after delimitating patches with code comments                                                         \\ \bottomrule
\end{tabular}
\vspace{-0.4cm}
\end{table}

\textbf{\stud.} In this study, the participants had to fix four bugs in a fixed time of 40 minutes. The used IDE was \textit{NetBeans} since all the students were familiar with this IDE---it was the reference IDE of the Java programming courses the students had attended before the SE course. When a bug was fixed, they had to delimit the corresponding patch (\ie the code they changed to fix the bug) by using code comments as follows: \texttt{/*Start <BUG REPORT ID>*/}, just before the patch, and \texttt{/*End <BUG REPORT ID>*/}, just after the patch. To share the bug reports and collect students' patches, a \textit{Google Forms}' questionnaire was~used. It is worth mentioning that technologies such as \textit{Git} and \textit{GitHub} are likely to be still unknown to students, especially if undergraduates. 

\textbf{Differences between \prof and \stud.}
In Table \ref{tab:diff}, we recap the differences in the experimental task and instrumentation between \prof and \stud. The professionals had 45 minutes to fix 6 bugs, with three interruptions, while the students had 40 minutes to fix 4 bugs, without any interruption.
It is worth mentioning that the students tackled a subset of the bugs tacked by the professionals (\ie the first four bugs), in the same order. The difficulty in fixing the two additional bugs assigned to the professionals was comparable to that of the first four bugs. Regarding the IDE, both professionals and students used an IDE they were familiar with. Indeed, for the students, the IDE was the one used in the Java programming courses. Finally, there were differences in the sharing of bug reports and collecting of patches since the professionals used technologies (\ie \textit{GitHub} and \textit{Git}) they usually dealt with.

\subsection{Metrics}
To measure the performance of professionals and students in fixing bugs, we used: $\boldsymbol{TaskCompletion}$ and $\boldsymbol{Speed}$. The former metric was computed as
\begin{math}
    {TaskCompletion}=\frac{\#BugsFixed} {\#Bugs} * 100
\end{math}, 
where $\#BugsFixed$ was the number of bugs the participant had fixed (correctly) during the bug fixing task on \textit{PacMan}, while $\#Bugs$ was the total number of bugs (\ie 6 for \prof and 4 for \stud). $TaskCompletion$ estimates performance since the time to accomplish the task was fixed (\ie 45 for \prof and 40 for \stud)~\cite{BergersenEtAllTSE2014}. The values of the \task metric range in [0\%, 100\%].  A value equal to 100\% indicates that a participant had fulfilled the task correctly and then his/her performance was the best possible.
As for the \textbf{\speed} metric, it was computed as \begin{math}
    {Speed}=\frac{\#BugsFixed} {TimeAllotted} * 60
\end{math}, 
where $TimeAllotted$ is the time allotted for the task. The values of the \speed variable range in [0, 8] for \prof and [0, 6] for \stud. Similar to $TaskCompletion$, ${Speed}$ estimates the performance construct. A value equal to 8 for ${Speed}$ indicates that the participant correctly fixed bugs with a speed of 8 bugs per hour. It is easy to grasp that the analysis for ${Speed}$ is slightly in favor of the professionals since they could theoretically achieve a higher speed (\ie up to 8 bugs per hour for \prof vs. 6 bugs per hour for \stud). The considered metrics are straightforward to facilitate a clearer understanding of the performance construct, and accurately represent it. In addition, our choice is well-founded given the specific nature of our contribution, and aligns with the objectives and context of our research.

\subsection{Data Analysis}
For each metric, we first computed some descriptive statistics and built boxplots, and then tested our null hypothesis, namely \textit{``there is no statistically significant difference between \stud and \prof with respect to the considered metric (\ie \task or \speed)''}. To that end, we used the Mann-Whitney U test since, whichever the metric was, the data were not normally distributed~\cite{Wohlin:2012}. We fixed
the significance level ($\alpha$) at 0.05. To estimate the magnitude of a statistically significant difference (if any), we used the Cliff's $\delta$ effect size. Finally, we deepened our analysis with a pairwise comparison between \stud and each site of \prof.

\section{Results}\label{sec:results}

\begin{figure}[t]
\includegraphics[height=4cm]{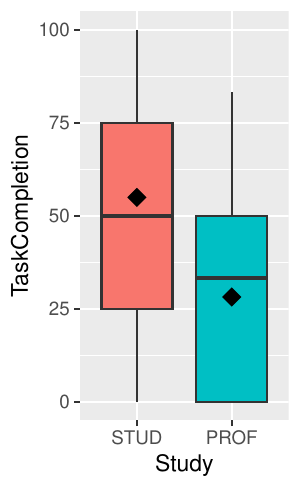}
\includegraphics[height=4cm]{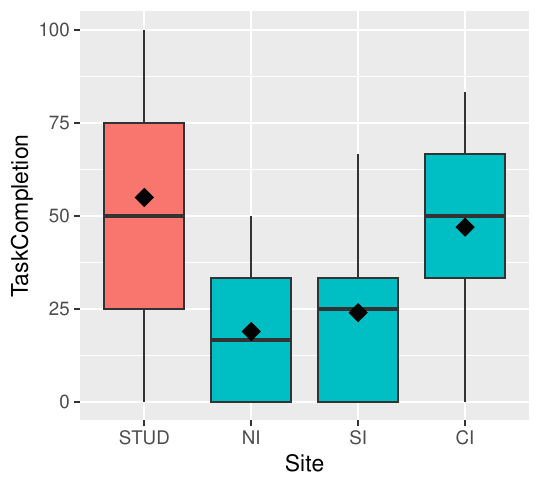}
\caption{Boxplots on the distributions for $\boldsymbol{TaskCompletion}$ by study and site---diamonds indicate the mean values.}\label{fig:taskcompletion}
\vspace{-0.3cm}
\end{figure}

\begin{figure}[t]
\includegraphics[height=4cm]{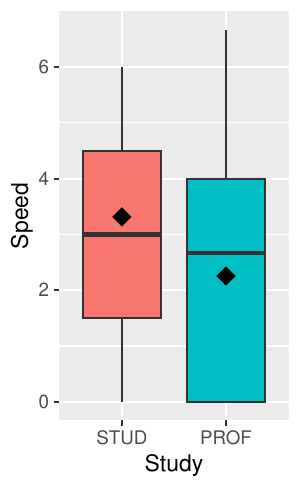}
\includegraphics[height=4cm]{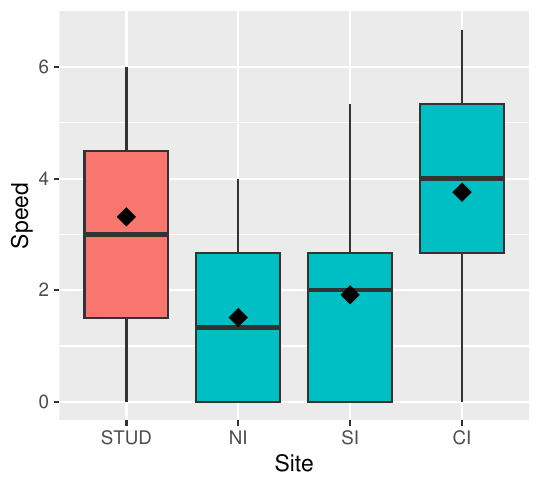}
\caption{Boxplots on the distributions for $\boldsymbol{Speed}$ by study and site--- diamonds indicate the mean values.}\label{fig:speed}
\vspace{-0.5cm}
\end{figure}

\hspace{\parindent}\ignorespaces \textbf{$\boldsymbol{TaskCompletion}$.} The boxplot in Figure~\ref{fig:taskcompletion} (left-hand side) suggests that the students had a better performance in fixing bugs than the professionals. Indeed, the box for \stud is higher than of \prof. The mean and median values are better for \stud (54.973\% and 50\%, respectively) than for \prof (28.175\% and 33.333\%, respectively). As for the hypothesis testing, the p-value returned by the Mann-Whitney U test was less than $\alpha = 0.05$ (first row in Table~\ref{tab:p-values}). Accordingly, we can reject the null hypothesis and accept the alternative one stating that \textit{``there is a statistically significant difference between \stud and \prof with respect to \task''}. This difference is surprisingly in favor of the students and its magnitude is medium (see Table \ref{tab:p-values}).\footnote{The Cliff's $\delta$ effect size is deemed: \textit{negligible}, if |$\delta| < 0.147$; \textit{small}, if $0.147 \le |\delta| < 0.33$; \textit{medium}, if $0.33 \le |\delta| < 0.474$; or \textit{large}, otherwise~\cite{Romano:2006}.} Finally, if we look at each site of \prof (see the right-hand side of Figure~\ref{fig:taskcompletion}), we can notice that the difference previously observed seems not to be consistent across all the sites: the performances of the students seem to be better than the one of the professionals involved in \torino and \napoli, but similar to the ones of the professionals involved in \roma. The results of the hypothesis testing confirm the existence of a statistically significant difference between \stud and \torino, and \stud and \napoli only (see the last three rows in Table \ref{tab:p-values}). These significant differences are confirmed even when adjusting the p-values, according to the Bonferroni correction, to counteract the multiple comparison problem. The magnitude of the differences is always large (Table~\ref{tab:p-values}).

\textbf{$\boldsymbol{Speed}$.} The boxplot in Figure~\ref{fig:speed} (right-hand side) suggests that the students outperformed the professionals on ${Speed}$. Based on mean and median values, the students fixed 3.315 bugs per hour on average and 3 on the median case, while the professionals fixed 2.254 bugs per hour on average and 2.667 on the median case. The p-value returned by the Mann-Whitney U test was less than $\alpha = 0.05$ (see the first row in Table~\ref{tab:p-values}), allowing us to reject the null hypothesis and accept the alternative one: \textit{``there is a statistically significant difference between \stud and \prof with respect to \speed''}. This difference in favor of the students and its magnitude is medium (Table \ref{tab:p-values}). Finally, the pairwise comparisons indicate the existence of a statistically significant difference between \stud and \torino, and \stud and \napoli only (last three rows in Table \ref{tab:p-values}), even after applying the Bonferroni correction. These differences were evident in Figure~\ref{fig:speed} (left-hand side) and in favor of the students. Their magnitude is always large, as Table~\ref{tab:p-values} shows. 

\begin{table}[t]
\caption{Hypothesis testing results.}
\label{tab:p-values}
\scriptsize
\begin{tabular}{@{}lllll@{}}
\toprule
\multirow{2}{*}{Comparison} & \multicolumn{2}{l}{\task} & \multicolumn{2}{l}{\speed} \\ \cmidrule(rl){2-3} \cmidrule(l){4-5}
& p-value  & Effect size & p-value & Effect size \\ \midrule
\stud vs. \prof & $<$0.001 & 0.471, medium & 0.001 & 0.368, medium\\ \addlinespace

\stud vs. \torino & $<$0.001 (0.001) & 0.606, large & 0.001 (0.002) & 0.565, large \\
\stud vs. \napoli & 0.001 (0.002) & 0.542, large & 0.003 (0.008) & 0.483, large \\
\stud vs. \roma & 0.33 (0.991) & - & 0.73 (1) & - \\ \bottomrule
\end{tabular}
\\\scriptsize{{*} In parenthesis, p-values adjusted according to the Bonferroni correction.}
\vspace{-0.5cm}
\end{table}

\textbf{Summary and Discussion.} The analysis for \task shows that the students outperformed the professionals when fixing bugs. This unexpected outcome is corroborated by the analysis of the \speed metric: despite the professionals theoretically achieving a higher speed, we still observed that their performance was worse than that of the students. We can postulate that this observed difference could be due to the different experimental settings: the professionals were exposed to stress factors.
However, the pairwise comparisons (on both the \task and \speed metrics) show that the difference in the performance of students and professionals is not consistent across all the sites of \prof. Specifically, we did not observe any significant difference between \stud and \roma, while we observed a significant difference for \torino and \napoli in favor of the students. Therefore, we can postulate that some characteristics of the participants (\eg in \roma, the participants were much younger in terms of seniority and characterized by a much higher number of master's graduates) and/or the working context could have made the difference. Rather than drawing definitive conclusions, our results feed the debate on the use of students in controlled experiments and pave the way for future investigations such as examining the complex factors influencing SE tasks. 

\section{Threats to Validity}\label{sec:limitation}
We discuss the threats that might affect the validity of our results based on the validity schema by Wohlin~\etal~\cite{Wohlin:2012}.

\textbf{Internal Validity.} Our study is potentially exposed to a \textit{selection} threat since the participation was voluntary for both the students and professionals and volunteers are generally more motivated than the whole population~\cite{Wohlin:2012}. 
A \textit{maturation} threat might affect our study since the students worked on \textit{PacMan} in the fifth experimental session (and after two training sessions). This might have positively influenced their performance. Namely, the students might have learned to fix bugs more efficiently over time. However, we do not deem this a major issue since the professionals were already skilled at fixing bugs---this was why they did not undergo any training in fixing bugs. Therefore, it is unlikely that such a threat has a significant impact when comparing students and professionals. 

\textbf{External Validity.} Although unlikely, we cannot exclude a threat of \textit{interaction of selection and treatment}. Namely, the participants in \prof and \stud might not be representative of the populations of software professionals and CS students, respectively. Finally, although in \prof, several design choices were taken to create a setting that more closely mirrored real-world industry conditions, there might still be a threat of \textit{interaction of setting and~treatment} since, for instance, seeded bugs were used.  

\textbf{Construct Validity.} We analyzed the data from a single experimental task. This potentially poses a threat of \textit{mono-operation bias}. Namely, our study might under-represent the (performance) construct and thus not give the full picture of the theory. 

\textbf{Conclusion Validity.}
We selected two samples from the target populations of professionals and students to study whether they perform similarly when fixing bugs. However, we did not take into account potential heterogeneity within these groups (except for comparing the single sites of \prof with \stud). Namely, we recognize a threat of \textit{random heterogeneity of subjects}. 

\section{Final Remarks} \label{sec:conclusion}
The research strategies that are mostly applied in SE are experiments, case studies, correlational studies, and surveys~\cite{StatusEmpiricalResearch}. Several studies report the rates of participation of students and professionals in experiments. H\"{o}fer and Tichy~\cite{StatusEmpiricalResearch} showed that 60\% of experimental studies in SE used students, whereas 22\% employed professionals, and only 14\% used both. Sj{\o}berg~\etal~\cite{Sjoberg:TSE2005} provided even higher rates regarding the participation of students in experiments: 87\% of the selected articles used students (only 9\% employed professionals). Moreover, undergraduates are used more often than graduates. Apart from the generalizability of results (\ie external validity issues), very few studies have been specifically carried out to compare the performance of students and professionals in their respective environments~\cite{Host:2000,Runeson2003,Salman:2015,PorterVotta1998,Counsell:2008}. Similar to these papers, we compare professionals and students, while differently from them, our research uniquely focuses on a task that professionals themselves have identified as particularly stressful compared to other kinds of tasks~\cite{RomanoCGVCSPROFES2024}. By selecting a task that mirrors the real pressures encountered in practice, we enhance the practical relevance of our study and underscore the value of investigating performance under conditions that truly matter to the industry. A key finding of our study is that professionals did not outperform students, challenging conventional wisdom. This unexpected difference from previous experiments challenges long-held assumptions and adds to the debate on their validity. Our findings encourage the community to examine the complex factors influencing SE tasks paying attention, for example, to make the setting of experiments as more realistic as possible. Also considering past research~\cite{Sjoberg:TSE2005,Sjoberg:ESEM2002}, we can argue that reproducing realistic settings (mundane realism) is more crucial than simply involving professionals. Therefore, the balance between realism and experimental control plays a key role in ensuring meaningful results that can transferred to the industry.

\section*{Data Availability}
The raw data is publicly available in our replication package on Figshare (\url{https://doi.org/10.6084/m9.figshare.28423796}).

\section*{Acknowledgements}
This work has been, in part, funded by the European Union - Next Generation EU, Mission 4, Component 1, under the PRIN program of the Italian Ministry of Universities and Research, project entitled “MOOD–
Mindfulness fOr sOftware Developers” (ID: D53D23008880006).
\balance

\bibliographystyle{ACM-Reference-Format}
\bibliography{bibliography}


\end{document}